# The DeMaDs Open Source Modeling Framework for Power System Malfunction Detection

David Fellner  
*Center for Energy*  
*AIT Austrian Institute of Technology*  
Vienna, Austria  
David.Fellner@ait.ac.at

Thomas I. Strasser  
*Center for Energy, AIT Austrian*  
*Institute of Technology* and *TU Wien*  
Vienna, Austria  
Thomas.I.Strasser@ieee.org

Wolfgang Kastner  
*Institute of Computer Engineering*  
*TU Wien*  
Vienna, Austria  
Wolfgang.Kastner@tuwien.ac.at

*Abstract*—Modeling and simulation of electrical power systems are becoming increasingly important approaches for the development and operation of novel smart grid functionalities – especially with regard to data-driven applications as data of certain operational states or misconfigurations can be next to impossible to obtain. The DeMaDs framework allows for the simulation and modeling of electric power grids and malfunctions therein. Furthermore, it serves as a testbed to assess the applicability of various data-driven malfunction detection methods. These include data mining techniques, traditional machine learning approaches as well as deep learning methods. The framework's capabilities and functionality are laid out here, as well as explained by the means of an illustrative example.

## I. Introduction

The development of new smart grid capabilities for electric power grids is essential these days. The transformation towards a sustainable, yet still resilient energy system entails various challenges. These demands can only be faced by novel functionalities [1], which allow the grid to react to the current situation. In order to implement them, but also to test and monitor them, realistic testbeds are needed. However, there are various obstacles to using the electrical power grid as a testbed. The reasons for this are mainly domain-specific: as the power grid is a vital building block of modern life, it is regarded as a critical infrastructure. Any meddling or introduction of non-fully elaborate functionality could compromise its reliability [2]. Moreover, the power grid can not be rebuilt in a scaled-down version that would fully reflect its properties. Furthermore, due to the historical development of the power grid as a hierarchical system, the lower tiers of the network are fairly ill-equip with sensors [3]. These circumstances make data collection and testing in the field, or on a replica of the actual power grid, either difficult or next to impossible.

This leads to modeling and simulation being the only feasible option for early-stage development and assessment of smart grid solutions. This is especially true if these approaches are not only to be tested in a very limited lab setting.

This work received funding from the Austrian Research Promotion Agency (FFG) under the "Research Partnerships – Industrial PhD Program" in DeMaDs (FFG No. 879017).

Regarding grid models, there is free material available to facilitate these tasks. Very prominent representatives there are the IEEE radial test feeders [4] which are widely used in power system analysis under novel circumstances [5]. Even though the IEEE test feeders feature load profiles, they lack renewable generation profiles and an approach for future scenarios in general. The SIMBENCH project [6] is an open-source project providing specifically designed power grids that allow for the simulation of distribution grids. These models also include scenarios and consumption or generation profiles for electric mobility, battery storage, and novel forms of power generation. In combination with load flow solvers or power grid simulation software [7], these resources can be used to assess the impact and behavior of new techniques in grid operation. The state-of-the-art on these solvers and tools is quite advanced [8] and allows for high computational efficiency [9]. The data generated in the course of this could also be used to develop means of monitoring grid-connected devices.

However, the integration of these solvers with grid simulation and the modeling of specific applications as well as their malfunctions is missing from the literature. This is a prerequisite for the development of monitoring applications. The current approaches are often solely mathematical models not integrating data-driven approaches [10]. In case they do integrate approaches such as machine learning, they only target very common issues and applications; in [11] the authors present a model for predicting general power consumption. The work presented in [12] is more specific focusing on combined heat and power as well as electrical vehicle integration into the power grid. Demand response in a smart grid environment is under scrutiny in [13], however, with mere attention paid to its implementation and not to its monitoring functionalities with regard to correct execution. When it comes to monitoring, significant contributions can be found in the field of security with respect to malicious attacks on the power grid [14]. Nevertheless, this does for example not cover misconfigurations occurring during regular operation. These misconfigurations can lead to malfunctions of the grid-connected device.

The framework presented now aims to fill this gap by providing modeling and data generation, processing, and analysis capabilities. It is designed to serve as a testbed aimed to develop and assess functional monitoring solutions. The approach strives to detect malfunctions during the regular operation of grid-connected devices. The grid setups to be used can be arbitrary. Also, the malfunctions under scrutiny can be modeled freely, as well as a variety of detection methods employed. This is demonstrated in detail in the previous works of [15] and [16]. These features allow for the easy expansion of monitoring use cases. Furthermore, the final detection application can be parameterized freely to facilitate development.

The manuscript has the following content: In Section I, the general motivation and background for the work the and field of application of the software framework are presented. Section II provides an overview of the framework, its architecture, and its functionalities. Section III provides insights into the application of the framework by illustrating an example use case in detail. Section IV outlines the impact the framework has as a testbed for the development of monitoring solutions for power system operators. Finally, Section V provides the conclusions and an outlook about potential further work.

## II. FRAMEWORK DESCRIPTION

The framework is entirely written in Python and the implementation can be found on the corresponding GitHub repository[1]. The most important dependencies regarding external libraries and their use in the framework are illustrated in Figure 1; almost all libraries used are free and open-source libraries, with the exception of a library to interface the here-employed power grid simulation software, DIgSILENT PowerFactory. As there is sample data provided in the repository, the use of such software is not mandatory. Furthermore, any grid modeling and simulation solution can be used in combination with the rest of the framework. In addition, a script which is under development is used for load estimation.

However, other implementations of this functionality can be used as well. This means there are no crucial parts of the framework that are not openly accessible. The common Python libraries are made for data handling and path allocations, whereas for the classic machine learning capabilities Scikit-learn [17] is used. For deep learning, especially for the recurrent neural networks employed, Pytorch [18] is being used. For regular neural network applications, Tensorflow [19] is applied. The choice of using different libraries for the implementation of artificial neural networks depending on their type was made in order to allow for increased flexibility when developing a solution. Pytorch enables the developer to adjust and craft the desired architecture in greater detail in comparison to TensorFlow. This is especially interesting when trying to craft a monitoring solution in a setting like the power grid, as the relevant properties of the data and features are widely unknown beforehand.

[1]https://github.com/DavidFellner/Malfunctions-in-LV-grid-dataset

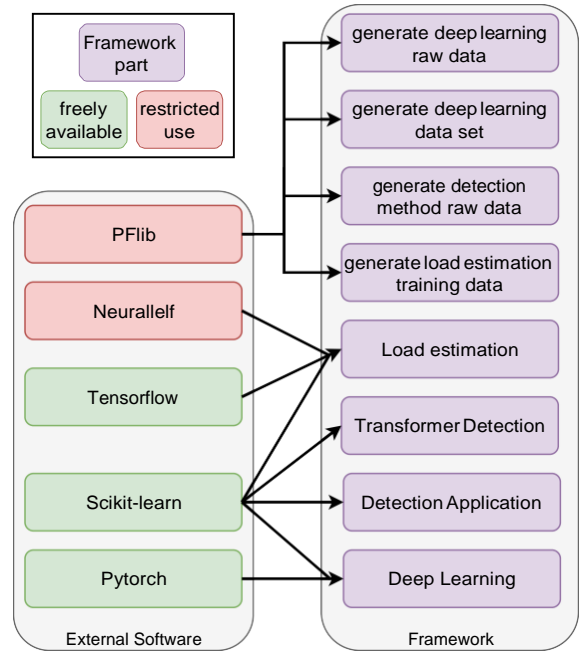

Fig. 1. The dependencies of the framework.

### A. Software Architecture

The architecture of the framework differs depending on the use case of the respective part of the software. Figure 2 depicts the software architecture of the framework.

The basic settings for the experiment to be conducted by the framework are defined in the configuration file. These settings include data paths and directories as well as configurations for the machine learning, or deep learning approaches that are to be used. The settings also define the neural network models and classifiers to use and how many layers or what type of kernel they should be parameterized with. Also, settings for the loading or creation of data and the assembly of datasets can be specified. These include the specification of the grid models or malfunctions, in order to define the use case the detection is applied to. Further settings include the mapping of data to align real-world measurements with simulation results, for cases in which these two data sources are to be combined.

Then, the data set generation or import of the defined use case is done via functions. Functions are chosen here in order to allow for easier integration of different data sources or grid simulation tools. The functional interfaces are easier to adjust or exchange in comparison to an integration of these data handlers within classes. Depending on the use case, this data is then saved. In the case of deep learning, the created data sets are also saved as their compilation is more computationally expensive compared to the data sets used for other approaches.

For experiments testing not a single detection method but a pipeline of methods that form an approach to a practical detection application, load estimation is done via an external script. This script is still under development and therefore not fully integrated with the Detection Application class. This also allows for the use of alternative load estimation or generally

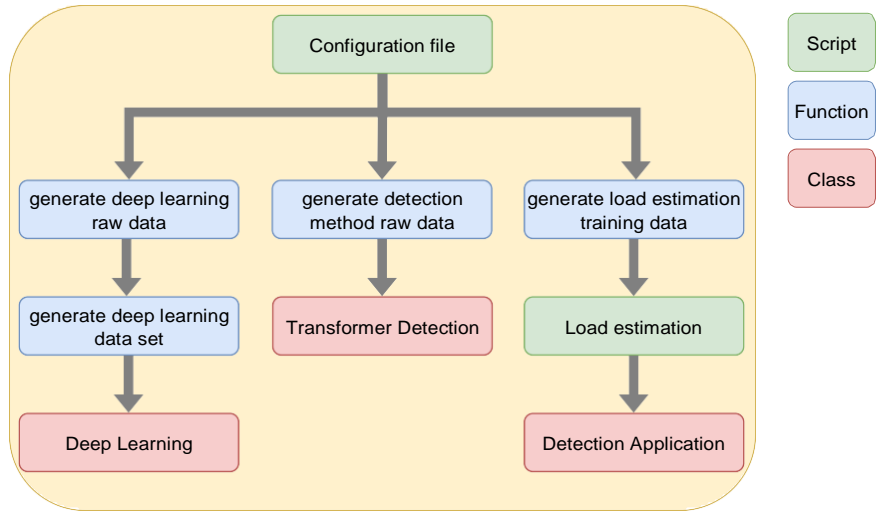

Fig. 2. The software architecture of the framework.

data mining approaches more easily.

The main functionalities regarding malfunction detection are bundled into three classes: Deep Learning and Transformer Detection both serve as isolated test beds for methods. These are either grid-unspecific using device-level data in the case of the Deep Learning approaches or grid-specific using transformer data for Transformer Detection. The last class, Detection Application, then allows for the integration of the individually assessed methods into a practically applicable detection application.

### B. Software Functionalities

The framework, as already mentioned above, allows for a great variety of scenarios in which the modeled malfunctions are to be detected. The malfunctions are modeled as incorrect control curves of devices whose behavior are reflected in grid operational data. This detection can be tested and validated in different grid topologies, using differently sized and composed data sets of different origins. The data can originate both from simulation or real-world settings such as lab environments. Furthermore, the approaches to preprocessing and data-driven detection can be varied. Also, various options for metrics and visualization of results are given.

When developing deep learning-based detection methods, data generation allows for the generation of large amounts of data. This is done by using an arbitrary number of grid models for simulation. These simulations can be parallelized, to swiftly yield operational data of a certain type of grid-connected device experiencing the malfunction modeled. Moreover, operational data of the correct behavior of these grid-connected devices is extracted as well. These data can not be obtained in the real world, especially not in a labelled manner as the occurrence of a misconfiguration goes unnoticed at the moment. The results are saved in a CSV format. This data is used to form data sets of the misconfiguration under scrutiny in the use case, which are stored in an hd5 format as they contain up to 200,000 samples. These data sets can now contain data stemming from a single grid or multiple grids. This allows for the assessment of whether the applied deep learning method is able to extract fundamental properties from the data. This is done in order to assess if a specific method can recognize a malfunction without any grid-specific context. The data and the individual samples therein can also be plotted. The framework allows for data preprocessing such as scaling as well as training in various deep-learning approaches. In addition, it enables a comparison to traditional statistical methods. Furthermore, hyperparameter tuning can be conducted. The performance is assessed using common measures such as the F-score, and scores can also be visualized.

Another monitoring approach is provided by transformer-level detection. Here only operational data gathered at the transformer is used. Data is loaded from, or generated and saved to CSV files. Also, both loading of, for example, real-world data, and generation of data is implemented to merge data of different origins. Then the data is preprocessed via Principal Component Analysis (PCA) and combined into datasets. Again, these datasets contain grid operational data of cases in which a malfunction is present or have their origin in regular grid operation. As there are data in a higher resolution as well as more data channels available in this setting, traditional machine learning approaches are to be tested here. This is due to the meters at substations measuring more variables, and these at a higher rate, than smart meters in the distribution grid. As this case is grid-specific, also more advanced tools of data analysis such as hierarchical clustering are available. This clustering helps to assess whether possible real-world data from a specific grid aligns with simulated data. Various classifiers can be applied which can then be assessed by the aforementioned range of result metrics and their plots.

The so-developed and assessed methods can be tested in a near-to-life setup which is represented by the detection application. Here, in order to fill gaps in data that were assumed to be known in the isolated method testbeds, also load estimation is conducted. A load estimation approach using a neural network is trained. Therefore, training data is generated in a similar manner to the cases described before and saved in a CSV file. Moreover, this load estimation is compared to a linear regression estimation to benchmark it. Using this estimation for data mining, data sets can be assembled in a manner similar to what they could also be collected like in the field. This aims at testing the performance of the detection methods under more realistic conditions. The data mining approach is also kept flexible in order to test the methods under different assumptions on which data is available. The result metrics can be inspected at every step of this pipeline to identify the potential for enhancements.

## III. ILLUSTRATIVE EXAMPLE

To complement the above-elaborated description of the software with a more tangible example, one use case is described in detail below (cf. Listing 1).

The crucial parts of a sample configuration file for testing a deep learning application on an electric vehicle charging station use case are presented. At first data paths are defined, both for the grid data used as well as for results and the dataset. Then the specific dataset to be used is defined along with the use case, which is done by choosing the device type that is to be monitored for malfunctions. Following, parameters for the type of neural network used for detection are specified along with training parameters such as the number of epochs, or the optimizer. The great flexibility in the choice of these parameters is made possible by the before-mentioned use of Pytorch. Also, the result metrics can be chosen, as well as settings for a grid search in order to be able to tune hyperparameters.

In the next section of the configuration file, the dataset to be created can be specified. If a dataset is already set to be available no new dataset is created. If not so, the number of samples the dataset created should contain, or how long a sample is, is defined. Also, the number of grids the samples should be drawn from can be specified. Lastly, settings on the grid simulation which creates the dataset can be customized. Parameters such as step size or how many cores should be used for parallelization can be set, along with the exact type of malfunction. In this case, as shown in Figure 3, a generic active power control curve of an electric vehicle charging station is inverted, which is considered the misconfiguration to be detected. The curve depends on the voltage, meaning in the malfunctioning case active power consumption is not reduced at low voltages which therefore constituted the detectable anomalous behavior. The red line marks the correct control curve, whereas the blue line is the inverted, malfunctioning control curve.

These settings and parameters are then used to either create or import a grid model. Such a grid model is depicted in Figure 4. The grid is modeled with the specified amount of, for example, photovoltaic units or electric vehicle charging stations. Some of them are then in turn modeled with the malfunction specified. Then grid simulations are run and data

Listing 1. Configurations for a deep learning use case.

```python
import os
import math

# Sytem settings
grid_data_folder = os.path.join(os.getcwd(),
 'raw_data_generation', 'input')
raw_data_folder = os.path.join(os.getcwd(),
 'raw_data')
...

# Deep learning settings
learning_config = {
    "mode": "train", # train, eval
    "dataset": "7day_200k",
    "type": "EV",
    # PV, EV, (PV, EV) > malfunction
    "RNN model settings": [1, 2, 20, 5],
    # dim of in&output, dim of hidden state, # of
     layers
    "LSTM model settings": [1, 2, 3, 5],
    "R-Transformer model settings": [1, 3, 2, 1,
     'GRU', 7, 4, 1, 0.1, 0.1],
    # input size, dimension of model,output size,
     heads, rnn_type, key_size, # local RNN
     layers, # RNN-multihead-attention blocks,
     dropout, emb_dropout
    "number of epochs": 20,
    "learning rate": 1 * 10 ** -6,
    "decision criteria": 'majority vote',
    ...
    "activation function": 'relu', # relu, tanh
    "mini batch size": 60,
    "optimizer": 'SGD', # Adam, SGD
    "k folds": 5, # choose 1 to not do crossval
    "early stopping": True,
    "LR adjustment": 'warm up',
    "% of epochs for warm up": 10,
    "train test split": 0.3,
    "metrics": ['accuracy', 'precision_macro',
      'recall_macro', 'f1_macro'],
    ...
    "plot samples": True,
    "classifier": "RNN",
    "save_model": True,
    "do grid search": True,
    "grid search": ("calibration rate", [0, 0.05,
     0.1, 0.2, 0.3, 0.4, 0.5, 0.6, 0.7, 0.8, 0.9,
     1])
}

# Dataset settings
raw_data_available = True # leave True if grid
 simulation is not available
sample_length = 7 * 96 # 96 datapoints per day
number_of_samples = 200000
number_of_grids = len([i for i in
 os.listdir(grid_data_folder)])

# Grid simulation settings
parallel_computing = True
cores = 12
sim_length = 365 # simulation length in days
step_size = 15 # simulation step size in minutes
percentage = {'PV': 0,
              'EV': 25, 'BESS': 0,
              'HP': 0} # percentage of busses with
               active PVs etc...
broken_control_curve_choice = 2 # 1 = flat curve, 2
 = inversed curve
t_start = None # default(None): times inferred from
 profiles in data
t_end = None
```

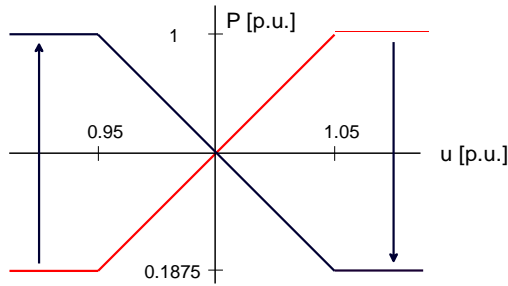

Fig. 3. Malfunctioning P(U) control curve.

is collected at the devices' connection points to the grids, which are symbolized by the triangles, boxes, or circles connected to the lines. The data is then used to assemble datasets. These are then used for training and testing the specified deep neural networks.

The so-trained neural networks are used for the detection of malfunctions in the test set. The performance results are then stored and also plotted, as Figure 5 illustrates. Here, the F-score is listed as a metric. The Precision, how accurate label predictions are, as well as the Recall, signifying how many of the true positives were found, are used to calculate this score. The results allow drawing conclusions about the performance of a certain parameterization of a certain deep neural network architecture on a specific dataset. It also allows for easy hyperparameter optimization. The model scoring the best results is saved and can be exported for integration into applications to make demonstrations easy. This should also help facilitate possible field tests of the found solution.

## IV. Impact and Application

The framework's impact is mainly threefold: first of all, it allows for the development of detection methods on a device level, as shown in the previous practical example. This method is intended to work across grid setups; the deep learning approach is meant to extract fundamental properties from the data of devices in regular operation and of devices experiencing malfunctions. Pretraining a network for a certain malfunction then allows the incorporation of the detection solution of this use case into a distribution system operator's monitoring system. Such a solution also enables the operator to know which malfunction occurred. The second aspect aims at developing a detection solution at the transformer level. This is done by using data collected at the substation and applying traditional machine learning methods to it. This detection approach is grid specific. However, it requires no extensive prior training. Only a certain calibration phase would be necessary.

For both application cases, different data sources, data qualities, and data availability can be assessed. Furthermore, different neural network architectures, classifiers, and parameters of these can be compared as well benchmarked against classic statistical methods.

Lastly, the full detection application merges the approaches mentioned above with a full detection application. This means integrating the isolated approaches with data mining techniques such as load estimation. This data mining is in turn also either performed by a neural network or by traditional statistical approaches. It can also be tuned to allow for optimal solution development for real-world applications. A testbed of this form did not exist to this point, and as elaborated in the beginning, the real-world power grid can not be used as such. Currently, because of the assumed data availability, its applicability is limited to the adaptation of misconfiguration detection in an LV grid segment linked to the MV level by a substation. However, for this reason, this scope of use cases also has a big advantage in integrability, since few alterations to the grid infrastructure are needed. Therefore, the framework has an impact as an enabler of technology development.

## V. Conclusions

The work presented describes the need for new monitoring capabilities for smart grids and points out the lack of possibilities to develop such with the means available. Therefore, a framework that can serve as a testbed for novel monitoring solutions for all sorts of new grid-connected devices is introduced here. Various approaches can be tested and integrated into a complete solution. This enables the development of a future detection tool for grid operators. The assessment of this solution can be conducted under as life-like circumstances as possible outside of the grid. The framework is designed in a flexible manner, as to allow users to exchange parts of it. Therefore, it is possible to use whichever means of grid simulation or data mining technique the user prefers.

In the future, more predefined use cases are to be added to reflect the characteristics of more malfunctions. Also, the choice and architectures of predefined machine learning algorithms ought to be updated regularly, in order to keep up with recent developments in these methods. Finally, a field test of the solution as a monitoring tool is envisioned.

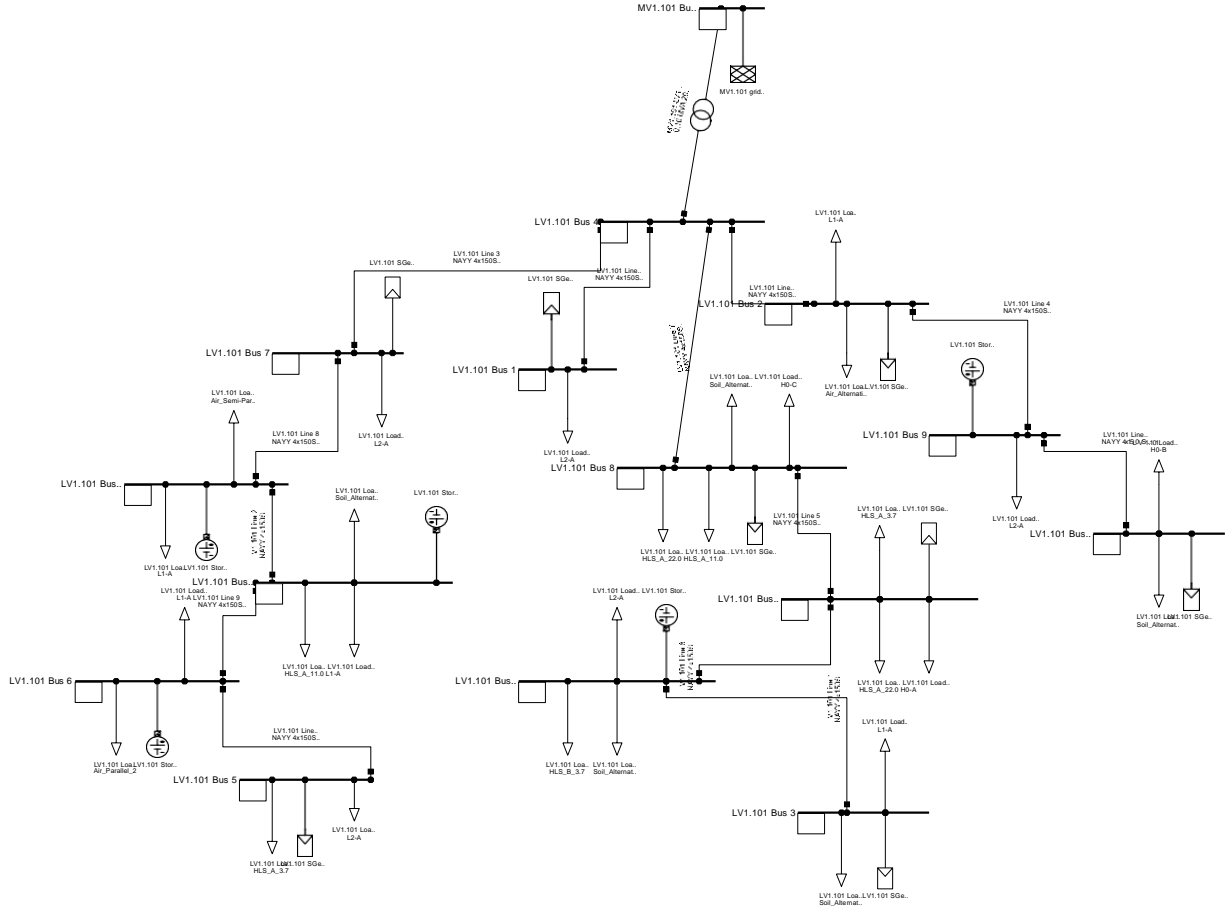

Fig. 4. Sample power grid (taken from [6]).

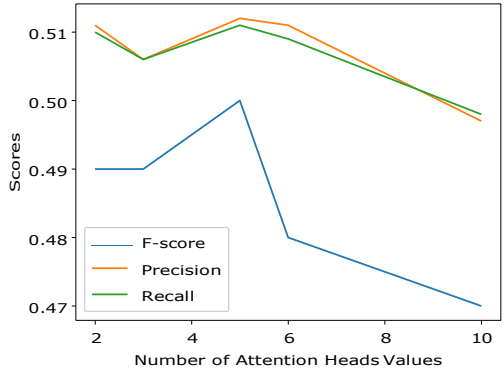

Fig. 5. Results on hyperparameter tuning.